\documentclass[a4paper]{article}

\usepackage{graphicx,color}

\usepackage{amsmath,amsfonts,amssymb,amsthm}

\newtheorem{defi}{Definition}
\newtheorem{teo}{Theorem}
\newtheorem{prop}[teo]{Proposition}
\newtheorem{cor}[teo]{Corollary}
\newtheorem{lem}[teo]{Lemma}
\newtheorem{rmk}{Remark}
\newtheorem{exm}{Example}

\begin{document}

\title{Generalizations of a method for constructing first integrals  of a class of natural Hamiltonians 
and some remarks about quantization}

\author{Claudia Chanu \\ \\
Dipartimento di Matematica e Applicazioni,\\ Universit\`a di Milano Bicocca.  Milano, via Cozzi 53, Italia.
\\
\\
 Luca Degiovanni, \, Giovanni Rastelli \\ \\ Formerly at Dipartimento di Matematica, \\ Universit\`a di Torino.  Torino, via Carlo Alberto 10, Italia.\\ \\ e-mail: claudia.chanu@unimib.it \\ luca.degiovanni@gmail.com \\ giorast.giorast@alice.it }
\maketitle

\begin{abstract}
In previous papers we determined necessary and sufficient conditions for the existence of a class of natural Hamiltonians with non-trivial first integrals of arbitrarily high degree in the momenta. Such Hamiltonians were characterized as (n+1)-dimensional extensions of n-dimensional Hamiltonians on constant-curvature (pseudo-)Riemannian manifolds Q. In this paper, we generalize that approach in various directions, we obtain an explicit expression for the first integrals, holding on the more general case of Hamiltonians on Poisson manifolds, and show how the construction of above is made possible by the existence on Q of particular conformal Killing tensors or, equivalently, particular conformal master symmetries of the geodesic equations. Finally, we consider the problem of Laplace-Beltrami quantization of these  first integrals when they are of second-degree.
\end{abstract}

\section{Introduction}

In recent years, several progresses have been done in the field of integrable and superintegrable Hamiltonian systems, both classical and quantum, by the introduction of 
new techniques for the study of higher-degree polynomial first integrals and higher-order symmetry operators. After researches exposed in 
\cite{CDR1}, \cite{KMK} and \cite{TTW}  is now possible to explicitly build and analyze Hamiltonian systems possessing symmetries of arbitrarily-high degree. 
For a more detailed introduction see the contribution to the QTS 7 proceedings written by W. Miller Jr. In several papers (\cite{CDR1}, \cite{sigma11}, \cite{POL}) we developed the analysis of a class of systems which, 
in dimension two, are a subset of the celebrated Tremblay-Turbiner-Winternitz (TTW) systems and are strictly related with the Jacobi-Calogero and Wolfes three-body 
systems \cite{CDR1}, \cite{POL}. In \cite{sigma11} we generalized these systems to higher-dimensions by introducing a $(n+1)$-dimensional extension $H$ of a given $n$-dimensional natural Hamiltonian $L$. We obtained necessary and sufficient conditions for the existence 
of a first integral of $H$ in a particular form, one necessary condition being the constant curvature of the configuration manifold on which $L$ is defined (for superintegrable systems with higher-degree first integrals on constant curvature manifolds see also \cite{MPY}). The first integral of $H$, which is independent from those of $L$, is polynomial in the momenta and can be explicitly constructed through a differential operator. In the present paper, we generalize the analysis done in \cite{sigma11} in several directions. 
In Sec.\ \ref{due} we extend the construction to non natural Hamiltonians on a general Poisson manifolds and obtain, also in this case, an explicit expression for the polynomial first integral. 
In Sec.\ \ref{tre} we restrict ourselves to cotangent bundles of (pseudo-)Riemannian manifolds and consider a wider class of higher-degree first integrals, 
we prove that a necessary condition for their existence is the presence of a particular class of conformal Killing tensors or, equivalently, of conformal master symmetries 
of the geodesic equations; we end the section with an example showing how the method can provide several independent first integrals of degree $m$. In Sec.\ \ref{quat} we 
characterize our construction in an invariant way and determine necessary and sufficient conditions for the constant  curvature or conformal flatness of the configuration 
manifold of $H$, conditions employed in Sec.\ \ref{cinq}, where the quantization of the second-degree first integrals obtained by our method is considered.

\section{Extensions on a Poisson manifold}\label{due}

Let us consider a Poisson manifold $M$ and a one-dimensional manifold $N$. For any Hamiltonian function $L\in\mathcal{F}(M)$ with Hamiltonian vector field $X_L$, we consider its extension on $\tilde{M}=T^*N\times M$ given by the Hamiltonian
\begin{equation}\label{HamExt}
H=\frac{1}{2} p_u^2+\alpha(u)L + \beta(u)
\end{equation}
where $(p_u,u)$ are canonical coordinates on $T^*N$ and $\alpha(u)\neq0$. The Hamiltonian flow of (\ref{HamExt}) is
\begin{equation*}
X_H=p_u\frac{\partial}{\partial u}-(\dot \alpha L+\dot \beta)\frac{\partial}{\partial p_u}+\alpha X_L,
\end{equation*}
where dots denotes total derivative w.r.t. the (single) variable $u$.

It is immediate to see that any first integral of $L$ is also a constant of motion of $H$, when considered as a function on $\tilde{M}$. We recall that a function $F$ is a first integral of $H$ if and only if $X_HF=\{H,F\}=0$.

In \cite{sigma11} we determined on $L$, $\alpha$ and $\beta$ necessary and sufficient conditions for the existence of two functions $\gamma\in\mathcal{F}(N)$ and $G\in\mathcal{F}(M)$ 
such that, given the differential operator 
\begin{equation}\label{U}
U=p_u+ \gamma(u) X_L,
\end{equation}
the function $U^m(G)$ obtained applying $m\neq0$ times $U$ to $G$  
is a non trivial additional first integral for $H$. 

In particular, if $L$ is a natural Hamiltonian on the cotangent bundle of a (pseudo-)Riemannian manifold $(Q,\mathbf{g})$
$$
L=\frac{1}{2}g^{ij}p_ip_j+V
$$
 and $\alpha$ is assumed to be not constant, an integral of the form $U^m(G)$ exists, with $G$ not dependent on the momenta, if and only if $G$ satisfy for some constant $c\neq 0$ the equations:
\begin{eqnarray}
&&\nabla_i\nabla_jG+mcg_{ij}G=0,\label{HessTeo}\\
&&\nabla^i V \nabla_iG=2mcVG,\label{VTeo}
\end{eqnarray}
which are equivalent to
$$
\{\nabla ^iGp_i,L\}=2mcGL,
$$
meaning that $\nabla^iGp_i$ is a conformal first integral of $L$.

If a solution of the previous equations exists, then the extended Hamiltonian (\ref{HamExt}) and the differential operator (\ref{U}) take the form
\begin{eqnarray}
\label{sH} H&=&\frac{1}{2}p_u^2+\frac{mc}{S^2_\kappa(cu+u_0)}L\\
\label{sU} U&=&p_u+\frac{1}{T_\kappa(cu+u_0)}X_L,
\end{eqnarray}

where the trigonometric tagged functions (see \cite{Ranada4,Ranada1}) are employed
$$
S_\kappa(x)=\left\{\begin{array}{ll}
\frac{\sin\sqrt{\kappa}x}{\sqrt{\kappa}} & \kappa>0 \\
x & \kappa=0 \\
\frac{\sinh\sqrt{|\kappa|}x}{\sqrt{|\kappa|}} & \kappa<0
\end{array}\right.
\qquad
T_\kappa(x)=\left\{\begin{array}{ll}
\frac{\tan\sqrt{\kappa}x}{\sqrt{\kappa}} & \kappa>0 \\
x & \kappa=0 \\
\frac{\tanh\sqrt{|\kappa|}x}{\sqrt{|\kappa|}} & \kappa<0
\end{array}\right.
$$
Here we show that an analogous result holds in a general situation.
\prop \label{pro2}
Let $H$ be the extension (\ref{HamExt}) of the Hamiltonian $L$ on the Poisson manifold $\tilde M$, let $U$ the differential operator $(\ref{U})$ and $G\in\mathcal{F}(M)$ a function such that $X_L(G)\neq0$. Then, $U^m(G)$ is a first integral for $H$ if and only if $G$ satisfies
\begin{equation}\label{EXL2}
X_L^2(G)+2m(cL+L_0)G=0 \qquad c,L_0\in\mathbb{R}.
\end{equation}
and $\alpha$, $\beta$ and $\gamma$ satisfy
\begin{eqnarray}
\label{Ea}
&& \alpha=-m\dot{\gamma},
\\
\label{Eb}
&& \beta=mL_0\gamma ^2 +\beta_0, \quad \beta_0 \in \mathbb R,
\\
\label{Ec}
&& \ddot{\gamma}+2c\gamma\dot{\gamma}=0.
\end{eqnarray}
\proof
In \cite{sigma11} it is proved that  
we have that $X_H U^m(G)=0$ for a function $G\in \mathcal{F}(M)$ if and only if  $L$, $\alpha$,  $\beta$ satisfy
\begin{eqnarray}\label{E1}
&& (m\dot \gamma+\alpha)X_L(G)=0,
\\
\label{E2}
&& \alpha\gamma X_L^2(G)-m(\dot \alpha L+\dot \beta)G=0.
\end{eqnarray}
Because $X_L(G)\neq0$, from (\ref{E1}) it follows that
\begin{equation}\label{Eac}
 \alpha=-m\dot \gamma
\end{equation}
and condition (\ref{E2}) becomes
\begin{equation*}
\dot \gamma \gamma \frac{X_L^2(G)}{G}=m\ddot{\gamma}L - \dot{\beta}.
\end{equation*}
Since $\dot \gamma=-\alpha/m\neq 0$, we get
\begin{equation}\label{Etemp}
\frac{X_L^2(G)}{G}=m\frac{\ddot{\gamma}}{\gamma\dot{\gamma}}L-\frac{\dot{\beta}}{\gamma\dot{\gamma}},
\end{equation}
which derived with respect to $u$ gives
$$
\frac{d}{du}\left(\frac{\ddot \gamma}{\gamma\dot{\gamma}}\right)L=\frac{d}{du}\left(\frac{\dot{\beta}}{m\gamma\dot{\gamma}}\right).
$$
But $L$ is a non-constant function on $M$, hence the functions $\ddot \gamma$ and $\dot\beta$ must be both proportional to $\gamma\dot{\gamma}$:
\begin{eqnarray*}
\ddot{\gamma}&=&-2c \gamma\dot{\gamma}=-c\frac d{du}\left(\gamma ^2\right),\\
\dot{\beta}&=&2mL_0\gamma\dot{\gamma} =mL_0\frac d{du}\left(\gamma ^2\right).
\end{eqnarray*}
By integrating and substituting in (\ref{Etemp}), we obtain conditions (\ref{EXL2}) and (\ref{Eb}).\qed
\begin{rmk} \rm
If $X_L(G)=0$ we trivially have $U^m(G)=p_u^m$, which is a first integral of $H$ only if $\alpha$ and $\beta$ are constant. Hence, it is a constant of motion functionally dependent on $L$ and $H$.
\end{rmk}
\begin{rmk} \rm
The equation (\ref{EXL2}) is obviously equivalent to 
$$\{L,\{L,G\}\}=-2m(cL+L_0)G;$$
this condition can be interpreted in terms of master symmetries: the Hamiltonian vector field $X_G$ is a master symmetry for the Hamiltonian vector field $X_L$ on 
the hypersurfaces $L=0$  or $G=0$. Further remarks about the special case when $L$ is a natural Hamiltonian are at the end of Sec.\ \ref{tre}.
\end{rmk}

By integrating the equations for $\alpha$, $\beta$ and $\gamma$ in Proposition \ref{pro2} the explicit expression for the extended Hamiltonian $H$ and the differential operator $U$ can be found.  From equation (\ref{EXL2}) we have \cite{sigma11}

\begin{teo} \label{teo} Let $H$ be the extension (\ref{HamExt}) of the Hamiltonian $L$ on the Poisson manifold $\tilde M$, let $U$ the differential operator $(\ref{U})$ and $G\in\mathcal{F}(M)$ a function satisfying $X_L(G)\neq0$ and (\ref{EXL2}). Then,  $U^mG$ is a first integral of $H$ if and only if $H$ and $U$ are in either one of the two following forms characterized by the value of $c$ in (\ref{EXL2})
\begin{enumerate}
\item  for $c\neq 0$  
\begin{eqnarray}
\label{Hc} H&=&\frac{1}{2}p_u^2+\frac{mc}{S^2_\kappa(cu+u_0)}(L+V_0)+W_0,\\
U&=&p_u+\frac{1}{T_\kappa(cu+u_0)}X_L,\nonumber
\end{eqnarray}
\item for $c= 0$ 
\begin{eqnarray}
\label{H0} H&=&\frac{1}{2}p_u^2+mA(L+V_0)+B(u+u_0)^2,\\
U&=&p_u-A(u+u_0)X_L,\nonumber
\end{eqnarray}
\end{enumerate}
with $\kappa,V_0,W_0\in \mathbb R$, $B=mL_0A^2$ and $A\neq0$.
\end{teo}

\proof By Proposition \ref{pro2}, $\alpha$, $\beta$, $\gamma$ must satisfy (\ref{Ea}), (\ref{Eb}), (\ref{Ec}). In the case $c\neq 0$ equation (\ref{Ec}) becomes
$\dot \gamma +c(\gamma ^2+\kappa)=0$, whose solution is 
$$
\gamma=\frac{1}{T_\kappa(cu+u_0)}.
$$
Hence, 
\begin{eqnarray*}
\alpha&=&\frac{mc}{S_\kappa^2(cu+u_0)},\\
\beta&=&\frac{ mcV_0}{S^2_\kappa(cu+u_0)} + W_0, 
\end{eqnarray*}
with $V_0=L_0/c$ and $W_0=\beta_0-m\kappa L_0$. In the case $c=0$, equation (\ref{Ec}) gives
$\dot \gamma+A =0$ with $A\neq 0$ in order to avoid $\alpha=0$. Hence,
\begin{eqnarray*}
\alpha&=&mA,\\
\beta&=&mAV_0+B(u+u_0)^2,\\ 
\gamma&=&-A(u+u_0),
\end{eqnarray*}
where $V_0$ is now an arbitrary constant.\qed

\begin{rmk} \rm
The constants $u_0$, $V_0$ and $W_0$ are not essential. Indeed, $H$ and $L$ are defined up to  additive constant $W_0$ and $V_0$ while  $u_0$ can be eliminated by a translation of $u$.
In the case $c \neq 0$, the choice $V_0=W_0=0$ gives the expressions (\ref{sH}) and (\ref{sU}) for $H$ and $U$ obtained in \cite{sigma11}. 
Moreover, by including  the constant $L_0$ in the Hamiltonian $L$, the condition (\ref{EXL2}) assumes the simpler form
\begin{equation*}
X_L^2(G)+2mcLG=0.
\end{equation*}
\end{rmk}

Once $L$ and $G$ satisfy condition (\ref{EXL2}) the first integrals $U^m(G)$ are explicitly determined for any $G: M\longrightarrow \mathbb R$. 

\teo \label{p8}
Under the hypothesis of Proposition \ref{pro2} the functions $U^mG$ can be explicitely written as
\begin{equation}\label{EE}
U^mG=P_mG+D_mX_LG,
\end{equation}
where
$$
P_m=\sum_{k=0}^{[m/2]}{\left( \begin{matrix} m \cr 2k \end{matrix} \right) \gamma^{2k}p_u^{m-2k}\left(-2m(cL+L_0)\right)^k},
$$
$$
D_m=\sum_{k=0}^{[m/2]-1}{\left( \begin{matrix} m \cr 2k+1 \end{matrix} \right) \gamma^{2k+1}p_u^{m-2k-1}\left(-2m(cL+L_0)\right)^k}, \quad m>1,
$$
where $[\cdot]$ denotes the integer part and $D_1=\gamma$.
\proof
From equation (\ref{EXL2}) it follows that for all $k\in \mathbb{N}$ we have 
\begin{equation}\label{fe1}
X_L^{2k+1}G=\left(-2m(cL+L_0)\right)^kX_LG, \quad X_L^{2k}G=\left(-2m(cL+L_0)\right)^kG.
\end{equation}
By expanding $U^m$ using  the binomial formula 
$$
U^mG=(p_u+\gamma X_L)^m=\sum_{k=0}^m\left(\begin{matrix} m \cr k \end{matrix}\right) p_u^k(\gamma X_L)^{m-k},
$$
and separating even and odd terms in $k$, by taking in account relations (\ref{fe1}) we get equation (\ref{EE}). 
\qed

The setting described in the previous section can be further generalized as follows.
Let $X_L$ be a Hamiltonian vector field on a Poisson manifold $\tilde M$, let on $\tilde M$
$$
X_H=Y+f_3X_L,
$$
for a vector field $Y$ and
$$
U=f_1+f_2X_L,
$$
where $f_i:\tilde M\rightarrow \mathbb R$. 
Following the same proof procedure as in \cite{sigma11} we get

\begin{prop} If $X_L(f_i)=0$ and $[Y,X_L]=0$ then $X_HU^m(G)=0$, i.e. $U^m(G)$ is a first integral of $H$, if and only if 
\begin{equation}
\left(f_1Y+(mY(f_2)+f_1f_3)X_L+f_2X_LY+f_2f_3X_L^2\right)(G)=-mY(f_1)G.
\end{equation}
\end{prop}

\proof
If $X_L(f_i)=0$ and $[Y,X_L]=0$,then
$$
\{H,L\}=0,
$$
$$
[X_H,U]=Y(f_1)+Y(f_2)X_L,
$$ 
$$
[[X_H,U],U]=0.
$$
Thus, 
\begin{eqnarray*}
 X_HU^m&=& U^{m-1}(m[X_H,U] + UX_H)=\\
&=&U^{m-1}\left(mY(f_1)+f_1Y+(mY(f_2)+f_1f_3)X_L+f_2X_LY+f_2f_3X_L^2\right).
\end{eqnarray*}
and the thesis follows. \qed

The analysis of such a generalization will not be considered here.

\section{Extensions of a natural Hamiltonian}\label{tre}
In the following sections we will assume that $L$ is a natural $n$-dimensional Hamiltonian on $M=T^*Q$ for a (pseudo-)riemannian manifold  $(Q, \mathbf{g})$:  
\begin{equation} \label{Lnat}
L=\frac 12 g^{ij}(q^h) p_ip_j +V(q^h),
\end{equation}
where $g^{ij}$ are the contravariant components of the metric tensor  and  $V$ a scalar potential.
This assumption, together with the hypothesis that $G$ is polynomial of degree $d$ in the momenta $(p_i)$, allows us to expand condition (\ref{EXL2}) into an equality 
of two polynomials in $(p_i)$ of degree $d+2$ that can be splitted into several differential conditions involving the metric, the potential and the coefficients of $G$.
Indeed, being $L$ a natural Hamiltonian, we have (in \cite{sigma11} the equation for $X_L^2$ was mistyped, however, this  does not affects any of the results of the paper,  
$$
X_L=p_i\nabla^i-\nabla_iV\frac{\partial}{\partial p_i},
$$
$$
X_L^2=p_ip_j\nabla^i\nabla^j-\nabla_iV\nabla^i-2p_j\nabla_iV\nabla^j\frac{\partial}{\partial p_i} - p_i\nabla^i\nabla_jV\frac{\partial}{\partial p_j}+\nabla_iV\nabla_jV\frac{\partial^2}{\partial p_i\partial p_j}.
$$

In \cite{sigma11} we dealt with the case $c\neq 0$, $d=0$, i.e. $G$ independent of momenta, obtaining the conditions (\ref{HessTeo}) and (\ref{VTeo}). The maximal dimension of the space of solutions of equation
(\ref{HessTeo}) is $n+1$ and it is achieved only if the metric $\mathbf{g}$ on $Q$ has constant curvature. We call {\it complete} the solutions $G$ of (\ref{HessTeo}) 
satisfying this integrability condition (see \cite{sigma11}).

In the following, we analyze in details the $d=1$ case ($G$ linear in the momenta), in order to show how the procedure works.

\begin{prop}\label{tGlin}
Let be  $G=\lambda^l(q^i)p_l+W(q^i)$. Then,  $U^mG$ is a first integral of $H$ if and only if
\begin{eqnarray}
&&\nabla^{(i}\nabla^j\lambda^{l)}+mcg^{(ij}\lambda^{l)}=0, \label{gl}\\
&&\nabla^i\nabla^jW+mcg^{ij}W=0, \label{Wg}\\
&&\nabla_iV(\nabla^i\lambda^l+2\nabla^l\lambda^i)+ \lambda^i\nabla^l\nabla_iV-2m\lambda^l(cV+L_0)=0, \label{Vl}\\
&&\nabla_iV\nabla^iW-2m(cV+L_0)W=0, \label{WV}
\end{eqnarray}
\end{prop}

\proof
For $G$ linear in the momenta we have
\begin{eqnarray*}
&&X_LG=p_ip_l\nabla^i\lambda^l-\nabla_iV\lambda^i+p_i\nabla^iW,
\\
&&X_L^2G=p_ip_jp_l\nabla^i\nabla^j\lambda^l-p_l(\nabla_iV(\nabla^i\lambda^l+2\nabla^l\lambda^i)
+ \lambda^i\nabla^l\nabla_iV)+\\ 
&&\hphantom{X_L^2G={}}+p_ip_j\nabla^i\nabla^jW-\nabla_iV\nabla^iW,	
\end{eqnarray*}
and condition (\ref{EXL2}) holds if and only if 
\begin{eqnarray*}
 p_ip_jp_l(\nabla^i\nabla^j\lambda^l+mc g^{ij}\lambda^l)+ p_ip_j( \nabla^i\nabla^jW+mc g^{ij}W)-\cr
p_l(\nabla_iV(\nabla^i\lambda^l+2\nabla^l\lambda^i)+ \lambda^i\nabla^l\nabla_iV - 2m\lambda^l(cV+L_0))+\cr
2m(cV+L_0)W - \nabla_iV\nabla^iW+=0,
\end{eqnarray*}
which is equivalent to eqs. (\ref{gl}, \ref{Wg}, \ref{Vl}, \ref{WV}).
\qed

\begin{rmk}\rm
The coefficients of terms with even and odd degree in the momenta  are involved in different equations: eq.s (\ref{Wg}) and (\ref{WV}) contain 
the ones of a $G$ independent of $p_i$. Hence, for $\lambda^i=0$ we recover the $d=0$ case: (\ref{Wg}) and (\ref{WV})  are the expansion in coordinates of 
(\ref{HessTeo}) and (\ref{VTeo})  for $G=W$. 
For $W\neq 0$ the compatible potentials $V$ have to satisfy both conditions (\ref{WV}) and (\ref{Vl}), thus it is impossible to get new potentials other than those 
compatible with a $G$ independent of the momenta i.e., satisfying conditions (\ref{Wg}--\ref{WV}). 
\end{rmk}
From (\ref{Wg}) one can derive (see \cite{sigma11}) integrability conditions for $W$ 
\begin{equation}\label{cint}
\left( R_{hijk}-mc(g_{hj}g_{ik}-g_{hk}g_{ij})\right)\nabla ^h \ln W=0.
\end{equation}
If these equations are identically satisfied we have complete integrability which is equivalent to constant curvature of $Q$, otherwise $W$ must satisfy all equations (\ref{Wg}) and (\ref{cint}). For example,
when $Q$ has dimension two, we have from (\ref{cint}) 
$$
\left(R_{1212}-mc\det (g_{ij})\right) \nabla^1 \ln W=0,
$$
and
$$
\left(R_{2121}-mc\det (g_{ij})\right) \nabla^2 \ln W=0.
$$
Therefore, because of the symmetries of the Riemann tensor, we have

\begin{teo}
If $Q$ has dimension $2$, then equations (\ref{Wg}) admit non-constant solutions $W$ only if $Q$ has constant curvature.
\end{teo}

For each $l$, the integrability conditions of (\ref{gl}) are weaker than those for the Hessian equation for $G(q^i)$ (\ref{HessTeo}) and therefore the curvature of 
$Q$ could be non-constant.

We give two examples in order to illustrate the Proposition \ref{tGlin}.

\begin{exm} \rm
As  shown in \cite{sigma11} and recalled above, when $Q$ has constant curvature, equation  (\ref{HessTeo}),  or equivalently equation (\ref{Wg}), admits a solution 
depending on  $n+1$ real parameters $(a_i)$. 
Let $G_i$ be  a solution determinated by the choice of a particular set of the $(a_i)$, let us assume that $G_i \neq G_j$. 
It is then natural to consider the relations between $U^mG_i$ and $U^mG_j$ and see if some choice of the parameters can provide new independent first integrals of the system. 
For example, 
let $L$ be the natural Hamiltonian on the constant curvature manifold $Q=\mathbb S^2$  with $(q^1=\theta,q^2=\phi)$
\begin{equation}\label{hsf}
L=\frac 12 (p_\theta^2+\frac 1{\sin ^2 \theta}p_\phi^2)+V.
\end{equation}
A complete solution of a 0th degree $G(\theta,\phi,a_1,a_2,a_3)$ has been computed in \cite{sigma11}
\begin{equation}\label{Ges}
G= (a_1\sin\phi+a_2 \cos\phi)\sin\theta+a_3\cos\theta.
\end{equation}
and for $a_3=0$, the integration of equation (\ref{VTeo}) -- or equivalently (\ref{WV}) -- gives
$$
V=\frac 1{\cos^2 \theta}F\left((a_1\sin \phi-a_2\cos\phi)\tan \theta\right).
$$
For different sets of the parameters $(a_k)$, $U^mG_i$ and $U^mG_j$ are no longer simultaneously first integrals of $H$ unless if $F=F_0=constant$, and therefore
\begin{equation}\label{ev}
V=\frac {F_0}{\cos ^2 \theta}.
\end{equation}
In this case, let be $G_1=G(a_1=1,a_2=0,a_3=0)$ and $G_2=G(a_1=0,a_2=1,a_3=0)$. Hence, for any extension of $L$ of the form (\ref{HamExt}) with  $\alpha$ given by (\ref{Ea}), 
the five functions
$L_0=L$, $L_1=p_2=p_\phi$, $H$, $U^mG_1$ and $U^mG_2$ are functionally independent first integrals of $H$. 
For $m=2$, recalling that $c=K/m$,   the curvature of $Q=\mathbb S^2$ is  $K=1$  and choosing for the other free  parameters of $\alpha$ the values $\kappa=0$ and $u_0=0$,
we have $\alpha=\frac {4}{u^2}$, and $U^2G_1$ and $U^2G_2$ are
\begin{eqnarray*}
U^2G_1&=&(\sin \phi \sin \theta )\left( p_u^2-p_\theta^2\frac 4{u^2}-F_0\frac 8{u^2\cos ^2 \theta}\right) +p_\theta p_u\frac 4u \cos \theta \sin \phi \\ 
&+&p_\phi p_u\frac 4u \frac {\cos \phi}{\sin \theta}-p_\phi^2\frac 4{u^2}\frac {\sin \phi}{\sin \theta},
\end{eqnarray*}
\begin{eqnarray*}
U^2G_2&=&\frac {(\cos ^4\theta+\sin ^2 \theta-\cos ^2 \theta)\cos \phi}{\sin^3 \theta}\left(p_u^2-p_\theta^2\frac 4{u^2} -F_0\frac 8{u^2\cos^2 \theta}\right)\\
&+&p_\theta p_u\frac 4u \cos \theta \cos \phi-p_\phi p_u\frac 4u \frac{\sin \phi}{\sin \theta}-p_\phi^2\frac 4{u^2}\frac{\cos \phi}{\sin \theta}.
\end{eqnarray*} 
\end{exm}

\begin{exm} \rm
We can use a complete solution $G(q^i,a_k)$ of  (\ref{HessTeo}) in order to construct
solutions $\lambda^i$ of (\ref{gl}). Namely, we can choose $\lambda^i=G_i$, $i=1, \ldots, n$ where $G_i$ 
denotes any particular solution of (\ref{HessTeo}). We remark that it is not necessary that $G_i\neq G_j$ for $i\neq j$, or $G_i\neq 0$ for all $i$. 
By substituting the $\lambda^i$ into (\ref{Vl}), the equations become $n$ second-order PDE in $V$ whose solutions provide examples of compatible potentials.
For instance, let us consider again $L$ given 
by (\ref{hsf}) on $Q=\mathbb S^2$.
We can choose for $\lambda^i$ the particular values $\lambda^1=\cos(\theta)$, $\lambda^2=0$ of (\ref{Ges})  as coefficients for a linear homogeneous $G$. 
Then, equations (\ref{Vl}) can be integrated yielding,
$$
V=\frac{c_1+c_2 \sin \theta}{\cos ^2 \theta},
$$
which, for $c_2\neq 0$ does not satisfies (\ref{VTeo}) with $G$ given by (\ref{Ges}), hence, for this potential the construction of $U^mG$ is possible only when $G$ depends on the momenta. 
For the different choice of $\lambda^i$,  $\lambda^1=0$, $\lambda^2=\cos \theta$,  the integration of (\ref{Vl}) gives
$$
V=\frac {c_1}{\sin^2 \theta},
$$
which is compatible with $G$ given by (\ref{Ges}) for $a_1=a_2=0$. The expressions of $U^mG$ can be computed by using (\ref{EE}).
\end{exm}

\begin{rmk} \rm
By considering the functions $\lambda^i$ as the components of a vector field $\mathbf \Lambda$, equation (\ref{gl}) can be written as
\begin{equation}\label{ms}
[\mathbf g,[\mathbf g,\mathbf \Lambda]]=-mc \mathbf \Lambda \odot \mathbf g,
\end{equation}
where $[\cdot,\cdot]$ are the Schouten-Nijenhuis brackets and $\odot$ denotes symmetrized tensor product. 
This means that $[\mathbf g,\mathbf \Lambda]$ is a particular kind of conformal Killing tensor, or, equivalently,  that $\mathbf \Lambda$ is a particular 
conformal master symmetry of the geodesic equations, where the conformal factor is a constant multiple of $\mathbf \Lambda$,  instead of an arbitrary vector field. 
In a similar way, for $G$ polynomial in the momenta of degree $k$ with highest degree term given by $\lambda^{i_1\ldots i_k}p_{i_1}\ldots p_{i_k}$, 
it is straightforward to show that a necessary condition for $U^mG$ to be first integral of $H$ is still of the form (\ref{ms}), where now $\mathbf \Lambda$ is a $k$-tensor field.
In the 0-th degree case $G=W(q^i)$ eq. (\ref{ms}) becomes $[\mathbf g,\nabla W]=-mc W\, \mathbf g.$ 
\end{rmk}

\begin{defi}
We call {\em self-conformal} ({\em s-conformal} in short)  the $(k+1)$-order conformal Killing tensor field $[\mathbf g,\Lambda]$ such that 
$$
[\mathbf g,[\mathbf g,\mathbf \Lambda]]=C\mathbf g\odot \mathbf \Lambda,
$$
$C\in \mathbb R$, is satisfied. In this case,  the $k$-order tensor  $\mathbf \Lambda$ is said to be a {\em s-conformal master symmetry} of the geodesic equations of 
$\mathbf g$. 
\end{defi}

In the case of $C=0$, i.e. $c=0$, s-conformal Killing tensors and master symmetries become the usual  Killing tensors and master symmetries.
 
\begin{teo}
 Let $G$ be a $k$-degree polynomial of degree $k$ in the momenta. A necessary condition for $U^mG$ 
to be first integral of $H$ is that the tensor $\mathbf \Lambda$ of components $\lambda^{i_1\ldots i_k}$ given by the coefficients of 
the highest-degree term  of $G$ is a self-conformal master symmetry of the geodesic equations of $\mathbf g$ or, equivalently, that $[\mathbf g,\mathbf \Lambda]$ 
is a self-conformal tensor field of $\mathbf g$ with $C=-mc$.
\end{teo}

The existence of a complete solution introduced in \cite{sigma11} and recalled above can be restated as follows

\begin{cor}
Equation (\ref{HessTeo}) admits a complete solution $G=W(q^i)$  if and only if the dimension of the space of the s-conformal Killing vectors $\nabla G$, 
with $C=-mc$, is maximal and equal to $n+1$. 
\end{cor}


\section{Intrinsic characterisation of the extended Hamiltonians}\label{quat}
We show under which geometrical conditions a $(n+1)$-dimensional natural Hamiltonian  can be written as  the extension (\ref{Hc}) of a natural Hamiltonian   $L$. 
Let us consider a natural Hamiltonian 
\begin{equation}\label{Hgen}
H=\frac 12 \tilde g^{ab}p_ap_b+\tilde V
\end{equation}
on a $(n+1)$-dimensional  Riemannian manifold $(\tilde Q, \tilde{\mathbf g})$ and let $X$ be a conformal Killing vector of 
$\tilde{\mathbf g}$, that is a vector field satisfying 
$$
[X,\tilde{\mathbf g}]=\mathcal L_X \tilde{\mathbf g}= \phi \tilde{\mathbf g},
$$
where $\phi$ is a function on $\tilde Q$ and $[\cdot,\cdot]$  are the Schouten-Nijenhuis brackets. We denote by $X^\flat$ the corresponding 1-form obtained by lowering the indices by means of the metric tensor $ \tilde{\mathbf g}$. 

\teo \label{teo1}

If on $\tilde Q$ there exists a conformal Killing vector field $X$ with conformal factor $\phi$ such that
\begin{eqnarray}
dX^\flat \wedge X^\flat=0, \label{Xnorm}\\
d\phi\wedge X^\flat=0, \label{phi_u}\\
d\left\|X\right\| \wedge X^\flat=0, \label{Xconst}\\
X(\tilde V)=-\phi\tilde V, \label{Vok}\\
\tilde R(X)=kX, \quad k\in \mathbb{R}, \label{Xeig}
\end{eqnarray}
where $\tilde R$ is the Ricci tensor of the Riemannian manifold,
then, there exist on $\tilde Q$ coordinates $(u,q^i)$ such that 
$\partial_u$ coincides up to a rescaling with $X$ and the natural Hamiltonian $(\ref{Hgen})$ has the form (\ref{Hc}).
\proof
Condition (\ref{Xnorm}) means that $X$ is normal i.e., orthogonally integrable: locally there exists a foliation of $n$ dimensional diffeomorphic manifolds $Q$ such that
$T_PQ=X^{\perp}=\{v\in T\tilde Q | \tilde g(v,X)=0\}$; 
it follows that there exists a coordinate system $(q^0=u, q^i)$ for $i=1,\ldots,n$ such that $\partial_u$ is parallel to $X$ and the components $\tilde g^{0i}$ vanish for all $i=1,\ldots,n$. Furthermore, by (\ref{phi_u}) the conformal factor $\phi$  is constant on the leaves $Q$ ($v(\phi)=0$ for all $v\in X^{\perp}$); thus, $\phi$
depends only on $u$. By expanding the condition that $X= F(q^a)\partial_u$ is a conformal Killing vector
$$
\big\{\tfrac 12 \tilde g^{00}(q^a) p_u^2 + \tfrac 12 \tilde g^{ij}(q^a) p_ip_j,   F(q^a)p_u \big\}=\phi(u)\big( 
\tfrac 12 \tilde g^{00}(q^a) p_u^2 + \tfrac 12 \tilde g^{ij}(q^a) p_ip_j\big)
$$
we get the equations
\begin{eqnarray} 
\label{pri}
\tilde g^{00} (2\partial_u F - \phi )- F \partial_u \tilde g^{00}=0&
\\ \label{sec}
\tilde g^{hj} \partial_h  F=0& \qquad j=1,\ldots,n
 \\ \label{ter}
 \tilde g^{ij} \phi + F \partial_u \tilde g^{ij}=0& \qquad i,j=1,\ldots,n
\end{eqnarray}
By (\ref{sec}), we have $ F= F(u)$, hence due to (\ref{ter}) we get that $\partial_u \ln \tilde g^{ij}$ is a function of $u$, the same function  for all  $i,j$. Thus, without loss of generality we can assume 
$\tilde g^{ij}=g^{ij}(q^h)\alpha(u)$. Moreover, Eq. (\ref{pri}) implies  that, up to a rescaling of $u$, $\tilde g^{00}$ is independent of $u$. 
By imposing $X(\tilde V)=-\phi V$, we obtain
 $\partial_u\ln \tilde V=-\phi/{F}$, that means $\tilde V=\alpha(u)V(q^h)$, thus we get 
$$H=\frac 12 g^{00}(q^h) p_u^2 + \alpha(u)\bigg(\frac 12 g^{ij}(q^h) p_ip_j +V(q^h)\bigg).$$ 
Finally, condition (\ref{Xconst}) means that the norm of $X$ is constant on $Q$, that is $F(u)^2g^{00}(q^i)$ is independent of $(q^i)$. This shows that up to a rescaling and a change of sign of $H$ we can assume $g^{00}=1$ and in the coordinate system $(u, q^i)$ (\ref{Hgen}) has the required form (\ref{Hc}). 
By computing again the Poisson bracket, we get relations between $ \phi$, $\alpha$ and $F$: 
$\alpha=k( F)^{-2}$ and $\phi=2 \dot F$ with $k$ a real not vanishing constant.
When $X$ is a proper conformal Killing vector,  we can assume that $\alpha$ is proportional to $F(u)^{-2}$.
The covariant components of the Ricci tensor  of $\tilde Q$ are given in Lemma \ref{lm}, in particular we have for $i=1,\ldots,n$  
\begin{equation*}
\tilde R_{00}=n \frac { \ddot F}F, \qquad  
\tilde R_{0i}=0.
\end{equation*}
Hence, $X=F(u)\partial_u$ is an eigenvector of the Ricci tensor with eigenvalue $\rho=n \frac { \ddot F}F$, which is constant if and only if $F$ is proportional to  $S_\kappa(cu+u_0)$. 
\qed

\begin{rmk} \rm
If $\phi=0$ (i.e., $X$ is a Killing vector), then $\alpha$ and $ F$ are necessarily constant
and this gives the geodesic term of the case $c=0$, but  equation (\ref{Vok}) does not characterize the potential of the Hamiltonian (\ref{H0}). 
\end{rmk}

\begin{rmk} \rm
It is straightforward to check that for a Hamiltonian of the form $H=\tfrac 12 p_u^2 + F^{-2}(u)L$ with
$L$ a natural $n$-dimensional Hamiltonian, $X=F\partial_u$ is a CKV with conformal factor $\phi=2\dot F$   such that
$X(F^{-2}(u) V)=-\phi(F^{-2}(u) V)$. Hence, conditions of the above theorem are necessary for having an extended Hamiltonian of our form.
\end{rmk}

We want now to study the geometric properties of the 
metric $\tilde{\mathbf g}$ obtained by an extension of a metric $\mathbf g$, in particular when $\mathbf g$ is of constant curvature.

In the following, we assume $\alpha (u)=f^{-2}$ in order to simplify computations. In particular, $f$ is allowed to be pure imaginary.
\begin{lem} \label{lm}
Let $(g_{ij})$ be the components of a $n$-dimensional metric on $Q$ in the coordinates $(q^i)$. We consider the $(n+1)$-dimensional metric on $\tilde Q$ 
having components $\tilde{g}_{ab}$  ($a,b=0,\ldots n,\ i,j=1,\ldots n$) with respect to 
coordinates $(q^0=u,q^i)$  defined as follows
\begin{equation} \label{part}
\tilde{g}_{ab}=\left\{ \begin{array}{ll}
1 & \qquad a=b=0, \\
0      & \qquad a=0, b\neq 0, \\
f^2(u)g_{ij}(q^h) & \qquad a=i, \ b=j. \\
                       \end{array}
\right.
\end{equation}
Then, the relations between the covariant components of the Riemann tensors associated with $\tilde{\mathbf g}$ and $\mathbf g$ are for all $h,i,j,k=1,\ldots,n$
\begin{eqnarray}
 \label{r4}
&&\tilde R_{hjkl}=f^2 R_{hjkl}- \frac{\dot f^2} {f^2} (\tilde g_{hk}\tilde g_{jl}-\tilde g_{hl}\tilde g_{jk}),
\\
&& 
\tilde R_{0jkl}=0 \label{r3},
\\
&&\tilde R_{0j0l}=
 -\frac{\ddot{f}}{f} \tilde g_{jl}.\label{r2} 
\end{eqnarray}
Moreover, the covariant components of the Ricci tensors $R_{ij}$ and $\tilde R_{ab}$ of the two metrics are related, 
for all $h,i,j,k=1,\ldots,n$, by 
\begin{eqnarray}
\label{ric1}
&&\tilde R_{00}=n \frac {\ddot f}f, \\
&&\tilde R_{0i}=0,\\
\label{ric3} 
&&\tilde R_{ij}=R_{ij}+ \left( f\ddot{f}+(n-1)\dot f^2\right)f^{-2} \tilde g_{ij},
\end{eqnarray} 
and the relation between the Ricci scalars $R$ and $\tilde R$ is
\begin{equation} \label{rs}
\tilde R= \frac R{f^2} +n \frac{2f\ddot{f}+(n-1)\dot f^2}{f^2},
\end{equation} 
where $\dot f$ and $\ddot f$ denote the first and second derivative w.r.t. $u$ of $f(u)$.
\end{lem}

Expressions (\ref{r4}), (\ref{ric3}), and  (\ref{rs}) become simpler when $Q$ is of constant curvature, while the other formulas remain unchanged.

\begin{lem} \label{lm2}
Under the hypotheses of Lemma \ref{lm} with $n>1$, if $\mathbf g$ is a  metric of constant curvature $K$,  then the non zero covariant components of the Riemann tensor $\tilde R$ associated with $\tilde{\mathbf g}$ are, for all $h,i,j,k=1,\ldots,n$
\begin{eqnarray}
\label{r4cc}
\tilde R_{hjkl}=\frac{K- \dot f^2 }{f^2} \,(\tilde g_{hk}\tilde g_{jl}-\tilde g_{hl}\tilde g_{jk}),
\end{eqnarray}
Moreover, the covariant components of the Ricci tensor $\tilde R_{ij}$  and   the Ricci scalar  $\tilde R$ are, for $i,j=1,\ldots,n$,
\begin{eqnarray}
\label{ric3cc} 
\tilde R_{ij}= \left( f\ddot{f}+(n-1)(\dot f^2-K)\right)f^{-2} \tilde g_{ij},
\\
\tilde R= n \frac{2f\ddot{f}+(n-1)(\dot f^2-K)}{f^2}.
\end{eqnarray} 
\end{lem}

\begin{teo} \label{curva2} Let $(Q,\mathbf g)$ be a $n$-dimensional Riemannian manifold of constant curvature $K=mc$ and $(\tilde Q,\tilde{\mathbf g})$ the extended manifold with metric (\ref{part}), therefore
\begin{enumerate}

\item the metric $\tilde{\mathbf g}$ is of constant curvature if and only if either $n=1$ or $m=1$ or $K=c=\dot f=0$,

\item the metric $\tilde{\mathbf g}$ is conformally flat  if and only if either $n>2$ or $\tilde{\mathbf g}$ is of constant curvature.

\end{enumerate}

\end{teo}

\proof For $n=1$ the extended metric is, up to a rescaling of $q^1$,
$$
\tilde{g}_{ab}=\left(\begin{array}{cc} 1 & 0 \\ 0 & f^2 \end{array} \right),
$$
which is of constant curvature if and only if $\ddot f$ is proportional to $f$  which is true if   $f$ is any trigonometric tagged function. For $n\geq2$, due to the Bianchi identities, the metric is of constant curvature  if 
the ratios
$$
\tilde R_{abcd}/ (\tilde g_{ac}\tilde g_{bd}-\tilde g_{ad}\tilde g_{bc})
$$
are independent of $(a,b,c,d)$,
that is by  (\ref{r4cc}), (\ref{r3}), and (\ref{r2})
\begin{equation}\label{condf1}
\ddot f f+ K-\dot f^2=0, 
\end{equation}
which for $c\neq 0$, i.e. $f^2=\frac {S^2_\kappa(\frac Km \,u+u_0)}{K}$, becomes 
$$
\frac{K^2(m^2-1)}{m^2}=0,
$$
which holds only for $m=1$ or for   $K=c=0$, when $f$ is constant (see Theorem \ref{teo}) and (\ref{condf1}) holds.  

For $n=2$ the three-dimensional extended metric $\tilde{\mathbf g}$ is conformally flat if and only if  the Weyl-Schouten tensor
\begin{equation*}
\tilde R_{abc}=\tilde{\nabla} _c\tilde{R}_{ab}-\tilde{\nabla}\tilde{R}_{ac}+\frac 1{2n}\left( 
\tilde{g}_{ac}\tilde{\nabla}_b\tilde{R}-\tilde{g}_{ab}\tilde{\nabla} _c \tilde{R}\right),
\end{equation*}
where $\tilde{\nabla}$ denotes the covariant derivative w.r.t. $\tilde{\mathbf g}$, vanishes.
By applying the formulas derived in Lemma \ref{lm2} we have that the only non vanishing components of $\tilde R_{abc}$ are, for $i,k=1,2$,
 $$
 \tilde R_{i0k}=\frac{\dot f}{f^3}\tilde{g}_{ik}(\ddot f f+ K-\dot f^2),
 $$
which, as shown above, vanish only for $m=1$ or in the  case when $0=K=c$ and $f$ is constant. For $n>2$ the $(n+1)$-dimensional extended metric $\tilde{\mathbf g}$ is conformally flat if and only if  the Weyl tensor
\begin{eqnarray*}
\bar C_{abcd}&=&\tilde{R}_{abcd}+\frac 1{n-1}\left(\tilde{g}_{ac}\tilde{R}_{bd}-\tilde{g}_{ad}\tilde{R}_{bc}+
\tilde{g}_{bd}\tilde{R}_{ac}-\tilde{g}_{bc}\tilde{R}_{ad}\right)+\\ &+&\frac {\tilde{R}}{n(n-1)}\left(\tilde{g}_{ad}
\tilde{g}_{bc}-\tilde{g}_{ac}\tilde{g}_{bd}\right).
\end{eqnarray*}
vanishes and, by applying Lemma \ref{lm2}, this is true for all manifold $Q$ of constant curvature.
\qed

\section{Quantization}\label{cinq}
We consider here quantization for the case $m\leq 2$ only. 
For $m=1$, it is well known how to associate a first order symmetry operator with any constant of motion linear in the momenta. 
In \cite{BCRII} the quantization of quadratic in the momenta first integrals of natural Hamiltonian functions has been analyzed  and we recall here the results relevant for our case.

Let  $\hat H$ be  the Hamiltonian operator associated with the Hamiltonian $H=\frac 12 g^{ij}p_ip_j+V$, we have
$$
\hat H=-\frac {\hbar ^2} 2\nabla_i(g^{ij}\nabla_j)+V=-\frac {\hbar ^2} 2 \Delta +V,
$$
where $\Delta$ is the Laplace-Beltrami operator.
Let $T=\frac 12T^{ij}p_ip_j+V_T$ be a first integral of $H$, let 
\begin{equation}\label{qK}
\hat T=-\frac {\hbar^2} 2 \nabla_i(T^{ij}\nabla_j)+V_T.
\end{equation}
We have (Proposition 2.5 of \cite{BCRII})
\prop \label{p1}
Let  be $\{H,T\}=0$, then $[\hat H,\hat T]=0$ if and only if
\begin{equation}\label{Cc}
\delta C=\delta (TR-RT)=0,
\end{equation}
where $R$ is the Ricci tensor, $T$ and $R$ are considered as endomorphisms on vectors and one-forms and
$$
(\delta A)^{ij\ldots k}=\nabla _r A^{rij\ldots k},
$$
is the divergence operator for  skew-symmetric tensor fields $A$.


\rm

For our purposes we need to apply (\ref{Cc}) to the Ricci tensor of the extended metric and to the constant of the motion  $T=U^2G$. By assuming constant the curvature $K$ of $Q$, the components of $\tilde R_{ab}$ are given by inserting $f^2=\frac 1K S^2_\kappa (\frac Km \, u + u_0)$ or $f^2=\frac 1{mA}$ in Lemmas \ref{lm} and \ref{lm2};  the covariant components of the Ricci tensor are given respectively by 
\begin{eqnarray*}
\tilde R_{00}&=& -n\frac{\kappa K^2}{m^2}, \cr
\tilde R_{0i}&=&0,\cr
\tilde R_{ij}&=&\frac{K^2}{m^2}  \left( n\kappa +\frac{(n-1)(m^2-1)}{(T_k(\frac Km u+u_0))^2}\right)g^{ij},
\end{eqnarray*}
for $K\neq 0$ and $\tilde R_{ab}= 0$ for $K=0$. 

In order to make computations easier, we remark that for $A$, $B$ two-tensors on a Riemannian manifold $(\tilde Q,\tilde{\mathbf g})$ we have
\begin{equation}
(AB-BA)^a_c=A^a_bB^b_c-B^a_bA^b_c=A^{ad}B_{cd}-g^{ad}g_{ec}B_{db}A^{be}.
\end{equation}

\lem For any symmetric tensor $T^{ij}$  the (1,1) components of $C=T\tilde R-\tilde R T$, where $\tilde R$ is the Ricci tensor of $\tilde{\mathbf g}$, are
\begin{eqnarray*}
&&C^0_0=0,\\
&&C^i_0=T^{0i}W,\\
&&C^0_i=-\tilde g_{ij}T^{0j}W,\\
&&C^i_j=0,
\end{eqnarray*}
where 
\begin{equation}\label{Wq}
W=(n-1)\frac{\ddot f f- \dot f^2+K}{f^2}.
\end{equation}
\rm

\begin{rmk} \rm \label{r5}
We immediately  have that 
if $W=0$ then $C=0$ and, by Proposition \ref{p1} , $\{H,T\}=0$ implies $[\hat H,\hat T]=0$. 
However, by Theorem \ref{curva2}, $W=0$ if and only if either $n=1$, $m=1$ or $f$ is constant, i.e., if and only if  $\tilde {\mathbf g}$ is of  constant curvature.
\end{rmk}

\teo For $m=2$,  $\{H,T\}=0$ implies $[\hat H, \hat T]=0$ if and only if $\tilde{ \mathbf g}$ is of  constant curvature i.e., if and only if $n=1$ or $f$ is constant.
\rm
\proof If $K=0$, and therefore $c=0$ and $f$ is constant,  then $W=0$. Otherwise, when $K\neq 0$ and  $c\neq 0$,
by computing   $T=U^2G$ and by applying Proposition \ref{p8} we get
\begin{eqnarray*}
&&T^{00}=G,\\
&&T^{0i}=\gamma \nabla^iG,\\
&&T^{ij}=-\frac K2\gamma^2 G g^{ij},
\end{eqnarray*}
where $\gamma$ is given by 
$$
\gamma=(T_\kappa (\frac Km \, u + u_0))^{-1},
$$
as proved in Theorem \ref{teo}. A straightforward computation gives
\begin{eqnarray*} 
\delta C_0&=&\gamma W\left(g^{il}\partial^2_{il}G+\partial_lG(\partial_ig^{il}+g^{il}\partial_i\ln \sqrt{g})\right)=\\
&=&\gamma W \Delta G=-\gamma nK W  G, \\
\delta C_i&=&f\partial_i G\frac {d}{du}\left(\gamma f W\right),
\end{eqnarray*}
where $g=\det (g_{ij})$.
By inserting the expressions  of  $\gamma$  and of $f^2=\frac {S^2_\kappa (\frac Km \, u + u_0)}{K}$ 
we have   that there are no non-trivial ($G\neq const.$) solutions to $\delta C=0$ other than those such that $W=0$, that is, after Remark \ref{r5},  when $n=1$ or  $\tilde Q$ is of constant curvature. 
\qed

In a recent paper  \cite{He}, where particular conformally flat, non-constant curvature manifolds are considered, it is shown that even if the Laplace-Beltrami quantization of some first integrals of the Hamiltonian fails, their quantization is somehow made possible by considering the  conformal  Schr\"odinger operator  instead of the standard (Laplace-Beltrami) one. The conformal Schr\"odinger operator is related to the standard  one by an additional term proportional to the scalar curvature.  In Theorem \ref{curva2}, we proved that our extended Hamiltonians for $n>2$ have always conformally flat configuration manifolds, therefore, the method exposed in \cite{He} could be, at least in principle, applicable. 


If we denote by $\tilde{\Delta}$ the Laplace-Beltrami operator of $(\tilde Q,\tilde{\mathbf  g})$ and by $\Delta$ the Laplace-Beltrami operator of the constant curvature manifold $(Q,\mathbf g)$, a direct calculation shows that 
\begin{equation}
\tilde{\Delta} =\partial _u^2 +n \frac {\dot f}f\partial_u +\frac K{f^2}\Delta,
\end{equation}
and
$
[\tilde \Delta, \Delta]=0.
$
Therefore, being
$$
\hat H=-\frac {\hbar ^2}2(\partial _u^2 +n \frac {\dot f}f\partial_u)+\frac K{f^2}\hat L,
$$
with
$$
\hat L= -\frac {\hbar ^2}2 \Delta +V,
$$
we have
\prop $\hat L$ is a symmetry operator of $\hat H$:
$$
[\hat H, \hat L]=0.
$$

\rm

Since $\hat H$ and $\hat L$ have common eigenfunctions, from $\hat H\psi=\mu\psi$ and $\hat L\psi =\lambda\psi$ we obtain for the eigenfunction of $\hat H$ the following characterization

\prop The function $\psi(u,q^i)$ is an eigenfunction of $\hat H$ if and only if $\psi$ is an eigenfunction of $\hat L$ and
\begin{equation}
-\frac {\hbar ^2}2(\partial ^2_u \psi+n\frac {\dot f}f\partial _u \psi)+\left( \frac {K\lambda}{f^2}-\mu\right)\psi=0.
\end{equation}

\rm

\section*{Acknowledgements}
This research was partially supported (C.C.) by the European program ``Dote ricercatori"
(F.S.E. and Regione Lombardia).

\end{document}